\documentclass[%
 reprint,superscriptaddress,amsmath,amssymb,aps,prl]{revtex4-2}
 
\usepackage{hyperref}
\usepackage{graphicx}% Include figure files
\usepackage{dcolumn}% Align table columns on decimal point
\usepackage{bm}% bold math
\usepackage{color}
\begin{document}
\preprint{APS/123-QED}
\title{Chirped Bloch-Harmonic oscillations in a parametrically forced optical lattice}% Force line breaks with \\
\author{Usman Ali}
\affiliation{%
 Department of Physics, Paderborn University, Warburger Strasse 100, D-33098 Paderborn, Germany}
 %\altaffiliation[Also at ]{Paderborn University, Department of Physics, Warburger Strasse 100, D-33098 Paderborn, Germany}%Lines break automatically or can be forced with \\
\author{Martin Holthaus}

\affiliation{%
 Institut f\"ur Physik, Carl von Ossietzky Universit\"at, D-26111 Oldenburg, Germany %\textbackslash\textbackslash
}%

\author{Torsten Meier}%
 %\email{torsten.meier@uni-paderborn.de}
\affiliation{%
 Department of Physics, Paderborn University, Warburger Strasse 100, D-33098 Paderborn, Germany}

\date{\today}
       
\begin{abstract}
%super-Bloch oscillations appears in a periodic lattice 
%shaken around a mean value
%Super-Bloch oscillations are the outcome of a detuning between the frequency of Bloch oscillations and the driving frequency with which a periodic lattice is either shaken or modulated. We analyze the dynamics for a model system in which a detuning is internally generated due to the position-dependent force provided by a parabolic trap. The detuning varies with position, thus changes during the evolution and is also dependent upon the drive phase. Hence, significantly different dynamics are obtained as the lattice is shaken with a modulated parabolic potential upon different choices of initial drive phases. We provide accurate explanations for the different obtained oscillatory transport and spreading regimes by analyzing the spatio-temporal dynamics in real space and by visualizing the relative phase in the k-space dynamics. We also compare our numerical results to an approximate semiclassical analytical expression for the group velocity for a modulated constant force system and find good agreement for coherent oscillations but deviations for oscillations with spreading dynamics which altogether supports the interpretations of our findings.
The acceleration theorem for wavepacket propagation in periodic potentials disentangles the k-space dynamics and real-space dynamics. This is well known and understood for Bloch oscillations and super Bloch oscillations in the presence of position-independent forces. Here, we analyze the dynamics of a model system in which the k-space dynamics and the real-space dynamics are inextricably intertwined due to a position-dependent force which is provided by a parabolic trap. We demonstrate that this coupling gives rise to significantly modified and rich dynamics when the lattice is shaken by a modulated parabolic potential. The dynamics range from chirped Bloch-Harmonic oscillations to the asymmetric spreading oscillations. We analyze these findings by tracing the spatio-temporal dynamics in real space and by visualizing the relative phase in the k-space dynamics which leads to an accurate explanation of the obtained phenomena. We also compare our numerical results to a local acceleration model and obtain very good agreement for the case of coherent oscillations, however, deviations for oscillations with spreading dynamics which altogether supports the interpretations of our findings.
\end{abstract}

\maketitle

\par{Ultracold atomic gases in optical lattices are ideal model systems for quantum simulations \cite{1}. These systems allow for a flexible manipulation of the parameters, inter-atomic interactions, and lattice defects. Improved measurement techniques has opened doors to new interpretations, effects, and applications of varied phenomena in solid-state physics \cite{1}. Prominent examples which are largely explored in ultracold atomic systems are superfluidity \cite{2}, quantum magnetism \cite{3}, topological matter \cite{4}, Anderson localization \cite{5} and Bloch oscillations \cite{6,7,8}.} 
\par{In solid-state physics the acceleration theorem is a fundamental concept which describes the motion of electrons in a crystal under the influence of spatially-homogeneous electric fields. The theorem states that for an electronic wavepacket which is well localized in k-space and is constrained to a single band, the packet's center $\textbf{k}_c$ evolves according to ${\hbar\dot{\textbf{k}}_c(t)} = -e \textbf{E}(t)  $, where $-e$ is the electronic charge, when an electric field $\textbf{E}(t)$ is applied \cite{9,10,11,12}. The acceleration theorem combined with Jones-Zener expression of group velocity \cite{13}, $v_g(t) = \hbar^{-1} dE/dk|_{k_c(t)}$, devise a powerful method to describe the wavepacket motion within a semiclassical framework. Important examples where this method is proven to be highly effective are Bloch oscillations (BO) \cite{14,15,16,17,18,19}  and super Bloch oscillations (SBOs) \cite{20,21,22,23,24,25,26}. Considering the dispersion relation of a one-dimensional single-band tight-binding model with hopping amplitude $J$ and lattice period $d$, $E(k) = -2J \cos(kd)$, these dynamics can be summarized by
\begin{equation} \label{eq:1}
		\dot{{k_c}}(t) = -\frac{1}{\hbar}F(t)  \quad   ;    \quad \dot{{x}(}t) = \frac{2Jd}{\hbar} \sin(k_c(t)d) .
\end{equation}
For $F(t)=F_0+F_D \sin(\omega_D t)$ representing the time-dependent force exerted by the electric field, BOs with frequency $\omega_B =F_0d/\hbar$ are obtained for the case $F_D = 0$ and SBOs appear if $F_D \neq 0$ and the frequency of the oscillating field $\omega_D$ is slightly detuned from a rational multiple of $\omega_B$ \cite{10}. 

The key principle behind the separable solutions in these elementary examples is the necessity of a position-independent force. However, in modern experiments with ultracold atoms in optical lattices with parabolic confinement different situations naturally arise. Thus position-dependent forces can be realized which give rise to an intricate interplay between the k-space and real-space dynamics. Considering the parabolic confining trap of strength $S$ a local acceleration theorem ${\hbar\dot{{k}}_c(t)} = -S x  $ can be defined for wavepackets located slightly away from the center of trap where the force varies slowly between lattice wells. The resulting dynamics are quite similar to BOs although the evolving wavepacket now dephases quite rapidly, which leads to collapse and revivals of BOs \cite{27,28,29,30,31}. This shows that the semi-classical approach discussed above will fail to describe all these effects, nonetheless the frequency and amplitude can be calculated quite generally. The situation becomes even more interesting and complex when the trap potential \cite{32} or the periodic lattice \cite{33} is modulated in time that can give rise to long-range transport for which the local acceleration model is expected to collapse.}
\par{In this Letter, we analyze the wavepacket dynamics in a periodic potential that is subject to a parametrically modulated parabolic trap. Our analysis shows that the modulation lead to oscillatory transport on top of BOs, for which the predictions of a local acceleration theorem are fairly accurate. These dynamics we denote as chirped Bloch-Harmonic oscillations (CBHOs). First, we show dephased CBHOs, where the dephasing follows a different mechanism compared to the dephasing of BOs in static systems. Next, we demonstrate that the dephasing of CBHOs can be suppressed by an initial phase shift in the drive leading to long-lived coherent CBHOs. Conversely, at an opposite phase of the drive, we observe an asymmetric oscillatory spreading of the wavepacket. The spreading motion reveal a new kind of mixed dynamics, which exists in the combination of standard BOs and anharmonic BOs. Furthermore, we present an adapted solution to the local acceleration theorem, which highlights the usefulness of the local force assumption in the undriven system and shows good agreement with CBHOs in the driven scenario.}

\par
\begin{figure}[tb]
\hspace*{-0.3cm} 
\includegraphics[scale=0.28]{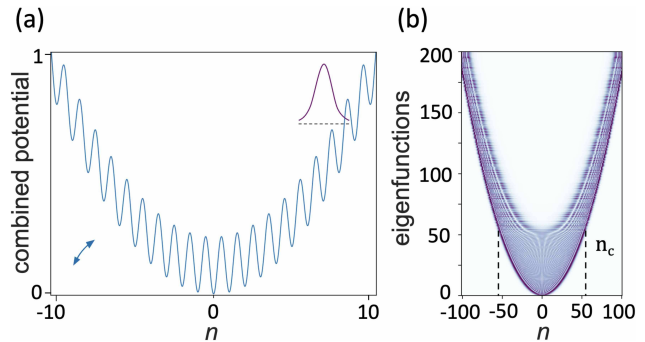}
\caption{\label{fig:epsart} (a) Schematic diagram showing the combined potential of a 1D optical lattice and an additional parabolic trap. The arrow indicates the oscillating curvature of the combined potential in the presence of a modulation of the trap and the inset represents a Gaussian wavepacket which we consider as the starting point of the quantum dynamics.
(b) Eigenfunctions of the time-independent system as a function of the real space index n. The dashed lines at $n=n_c$ mark the separation between harmonic oscillator like and spatially-localized eigenfunctions.}\label{fig:1}
\end{figure}

{We consider a condensate of ultracold rubidium atoms in an axially-symmetric crossed optical dipole trap which provides loose axial confinement, as compared to tight confinement along the transverse plane. In the limit of strong transverse confinement, and considering the atom-atom interactions are tuned to zero by using the Feshbach resonance, the effective potential in the axial direction is parabolic \cite{34}. In addition a one dimensional (1D) optical lattice is introduced alongside the parabolic potential, which results a symmetrically-curved periodic potential, see Fig.~\ref{fig:1}. We assume that the dynamics starts with a rapid displacement of the center of the parabolic potential which together with its subsequent time-dependent modulation can be realized by specialized optical modulators, such as acousto-optic modulator, in the axial beam \cite{35}. Thus the parabolic potential is modulated periodically which means that the overall curvature oscillates in time. In dipole and rotating wave approximations, the dynamics of ultracold atomic condensates in the combined potential of a 1D optical lattice and a modulated parabolic trap are effectively described by the Hamiltonian
%Such driving of a trap is considered in analytical studies \cite{32,33,34} and appears to be also experimentally feasible \cite{35}.
\begin{eqnarray}\label{eq:2}
 H=\frac{{p}^2}{2m}+\;V_{o}\;\sin^2{\left(\frac{\pi}{d}x\right)} \quad\quad\quad\quad\quad\quad\quad\\
 +\frac{1}{2}m\omega_{\tau}^2x^2\{1+ \alpha \sin(\omega_D t+\phi)\} , \;  \nonumber
 \end{eqnarray}
with $t>0$. Here, $V_o$ is the depth of the optical lattice, $d$ is the lattice period, $\omega_{\tau}$ is the frequency of the parabolic trap, and $m$ is the atomic mass. Moreover, $\alpha$, $\omega_D$, and $\phi$ denote amplitude, frequency, and initial phase of the driving field, i.e., the oscillatory part of the parabolic potential, respectively.}

\par{If the lattice depth is sufficiently high as compared to the atomic recoil energy $E_R= \hbar^2\pi^2/2md^2$, i.e. $V_0\gtrsim 4E_R$, and the modulation does not induce interband tunneling, one can express the wavefunction $\psi(x,t)$, as a superposition of Wannier functions $w(x)$ that are localized in the individual wells at $x=nd$ via $\psi(x,t)= \sum_n c_n(t) \;\!w_0(x-nd)$. In such a single-band tight-binding description of the system, the dynamics of the complex amplitudes $c_n(t)$ follow from the discrete time-dependent Schr\"odinger equation and reads
\begin{equation} \label{eq:3}
	i\hbar \; \frac{\partial {c}_n }{\partial t}	= -\frac{J}{2} (c_{n+1}+c_{n-1}) + {K(t)} \; n^2 c_n ,
\end{equation}
with $K(t) = K_0+K_D \sin(\omega_Dt+\phi)$, the parabolicity $K_0 = m\omega_{\tau}^2d^2/2$, and the modulated parabolicity $K_D = \alpha K$ \citep{27}. The symbol $J$ denotes the nearest-neighbor tunneling matrix element, which depends upon the scaled depth of the optical lattice, $s=V_o/E_R$, as $J/E_R \sim {8}\;{{\left(s\right)}^{\frac{3}{4}}}e^{-2\sqrt s}/{\sqrt\pi}\;$.}

\par{In order to initiate the quantum dynamics, we start with a sudden displacement of the parabolic trap center at time $t = 0$, along the axis of the combined potential. The displacement induces a shift in the mean position of the atomic cloud and the ensemble starts its journey over the curved periodic wells of the parabolic lattice at $x_0/d=n_0$.  For a displacement above the critical index, $n_c= (2J/K_0)^{1/2}$ the energy of the wells is larger than the trap free bandwidth $E>2J$. In such a regime the eigenfunctions are increasingly localized in space \cite{36,37}, as shown in Fig.~\ref{fig:1}(b). Without considering a modulation of the trap potential the expected dynamics would be quite similar to BOs in a locally static force of strength ${F_{n_0}} \sim 2Kn_0/d$ \cite{27}. We assume a preparation of the atomic condensate in the regime of localized eigenfunctions and describe it by a displaced Gaussian wavepacket
\begin{equation} \label{eq:4}
	c_n(t=0) = \frac{1}{\sqrt{\sigma_n \sqrt{\pi}}}\; e^{-\frac{(n-n_o)^2}{2\sigma_n^2}} e^{-ik_o n} ,	
\end{equation}
with mean, initial momentum, and the spatial-width represented by $n_0$, $k_0$, and $\sigma_n$, respectively. To obtain the dynamical evolution we solve Eq.~\eqref{eq:3} numerically by using the fourth-order Runge-Kutta method for the initial conditions given by Eq.~\eqref{eq:4}. To trace and to be able to analyze the wavepacket dynamics in k-space we evaluate the Fourier transform
\begin{equation}
		c_k(t) = \frac{1}{\sqrt{2\pi}} \int_{-\infty}^{\infty} c_n(t)\; e^{ikn} dn,
\end{equation}
with $k$ scaled in units of $k_B = 2 \pi /d$ and restricted to the first Brillouin zone, i.e., $-1/2 \leq k \leq1/2$.}
\par{We use values of the system's parameters from an experiment with Bose-condensed ${{}^{87}}$Rb atoms \cite{38}, with atomic mass $m=1.1443\times10^{-25}$~kg, which are placed in an optical lattice of period $d=397.5$~nm with depth $V_0 = 12.77~E_R$ and a parabolic trap with frequency $\omega_{\tau}=2\pi\times9$~Hz. These values corresponds to $J=0.024~E_R$, $K_0=1.52\times10^{-5}~E_R$, and $n_c\cong56$. The experimental procedure matches very closely with our assumption of the wavepacket preparation, with the only difference that we chose a parabolic optical trap, instead of a magnetic trap, and that in our model we consider that the trapping potential can be modulated periodically in time. We start the dynamics from $n_0=125$, such that the initial Bloch frequency is given by $ \hbar\omega_B=2K_0 n_0 = 0.0038~E_R$, which corresponds to $\omega_B=2\pi\times13.76$~Hz. A rather wide initial wavepacket with width in real space of $\sigma_n=3.16$ corresponding to distance $7.95~\mu m$ is considered, which is appropriate in the weak trapping used here. The trap potential is modulated with strength equal to the static trap strength, i.e., $\alpha =1$, and the modulation frequency is tuned to exactly match the initial Bloch frequency, i.e., $\omega_D=\omega_B$. %Due to the rather low value of the trapping frequency in comparison to other experiments, the parabolic trap potential can be considered as weak.
%Thus the distribution of atoms is on several optical lattice wells, and the overall peak lies at the minimum of a lattice well due to the Blue detuned optical field.
 As is shown below, different dynamics are obtained when tuning the initial phase $\phi$ of the time-dependent trap potential which therefore acts as a control parameter.}
% The initial width in quasimomentum space is determined by the relation $\sigma_n \, \sigma_k = 1/2$.}
% The dynamics of the average group velocity is obtained from the gradient of the mean position as function of time. For the undriven static system the eigenfunctions shown in Fig.~\ref{fig:1}(b) are obtained by the imaginary time propagation method applied to Eq.~\eqref{eq:3}.}
\begin{figure}[tb]
\hspace*{-0.3cm} 
\includegraphics[scale=0.44]{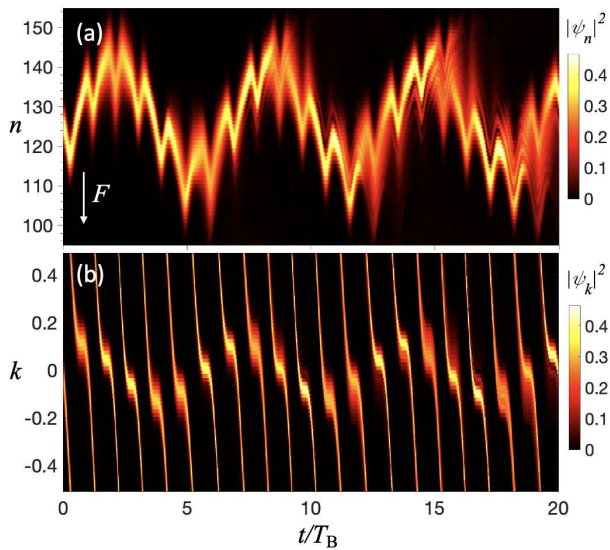}\caption{\label{fig:epsart} Time evolution of the absolute square of the wavefunction in (a) real and (b) quasi-momentum space
exhibiting CBHOs with some additional weak dephasing.
The wavepacket starts its journey at $t=0$ at $n_0=125$ with $k_0=0$ and we consider a modulated trap potential with drive phase $\phi = 0$.}\label{fig:2}
\end{figure}
\vspace{0.2cm}
\par{The dynamics obtained for our spatially-inhomogeneous and periodically-driven system for $\phi=0$ is shown in Fig.~\ref{fig:2}. As seen, the result demonstrate coupled evolution in real and quasi-momentum space where the wavepacket's center in one space moves in accordance with the other and the result correspond to CBHOs. A slow oscillatory transport combined with standard BOs is observed in Fig.~\ref{fig:2}(a), which completes its period in 6.5 Bloch periods moving across more than 50 lattice sites in real space. The transport carrying oscillation highlight the existence of an effective relative phase between the modulation and the BOs in the shaken periodic lattice even though an external detuning is absent. The dynamics resemble SBOs, however the corresponding evolution in k-space, shown in Fig.~\ref{fig:2}(b), reveals that the relative phase do not sweep around the whole BZ but rather oscillates around its center. The oscillating phase is an outcome of coupling between coordinate-position and quasi-momentum that manifests itself in the form of parametric modulation in Bloch frequency. In more detail, the relative phase develops from spatial variations, which increase (decrease) the Bloch frequency for wavepacket transport against (in) the direction of force. With this, the cycle averaged momentum changes its sign that flips the transport direction. Accordingly, Bloch frequency starts to decrease (increase) and the phase oscillates. Further, this mechanism repeats itself and the cycle continuous. As a result, quick oscillations are generated and the real space amplitude is low as compared to standard SBOs. Keeping in view the harmonic oscillator-like profile of oscillatory transport, it is noted that the period of CBHOs, i.e. the harmonic-period, is independent of wavepacket's initial position and is given by $T_{HO}\sim 2\pi\hbar/\sqrt{JK_0\alpha}$.}

\begin{figure}[t]
\includegraphics[scale=0.4]{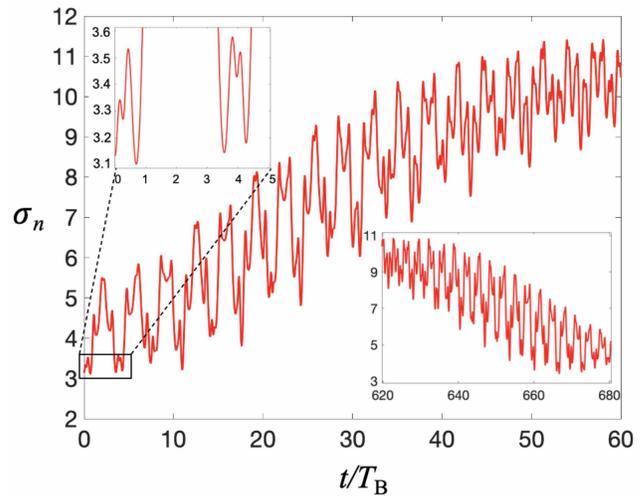}
\caption{\label{fig:epsart} Width of the wavepacket in real space for a drive phase of $\phi=0$. The left inset shows the magnified image of the width during the first five Bloch periods. The inset on the right displays the long time evolution of the width which shows a decline of the width corresponding to a complete revival. }\label{fig:3}
\end{figure}

\begin{figure}[t]
\includegraphics[scale=0.55]{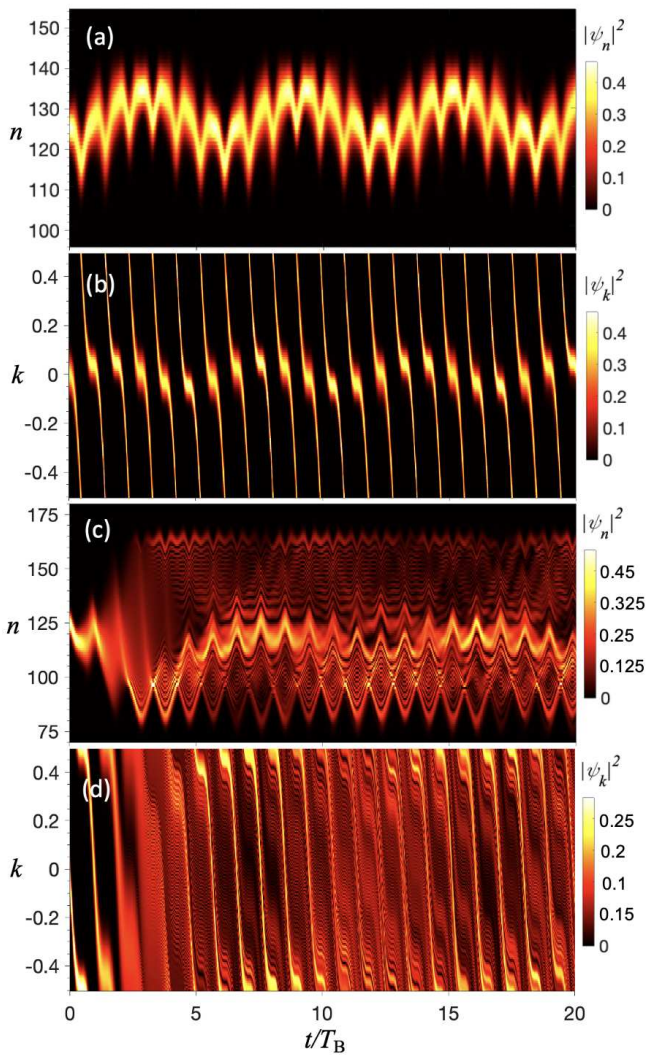}
\caption{\label{fig:epsart} Drive-phase-dependent dynamics. Time evolution of the absolute square of the real space wavefunction for a drive phase of (a) $\phi = -\pi/2$ and (c) $\phi = \pi/2$ exhibiting coherent CBHOs and asymmetric spreading dynamics, respectively. The corresponding quasi-momentum evolutions are shown in (b) and (d), respectively.} \label{fig:4}
\end{figure}
\par{The CBHOs shown above are accompanied by slow dephasing, which is initiated by a broadening of the wavepacket at times when the transport changes direction, as shown in Fig.\ref{fig:2}. Likewise, the wavepacket evolution in k-space also shows broadening, although the width in k-space is inverse to the width in real space. In the presence of broadening, a collapse of the coherent oscillations occurs, and the long time dynamics are heavily dephased. To analyze the dephasing, we plot the time evolution of the width in real space, i.e., the square root of the variance, in Fig.\ref{fig:3}. Starting with the first harmonic-period, we note that the width mainly increases when the wavepacket moves against the force and becomes smaller during the motion in the direction of the force. However, the initial width is never returned after the first Bloch period (see insets in Fig.\ref{fig:3}), and there is overall growth every harmonic-period. The width is reduced at the start of the second CBHO, which again increases in addition to a growth factor from the previous oscillation. This pattern is repeated until the width saturates at a maximum value. The oscillation of the width continues in the saturation region until an interval of coherent dynamics reappears. The inset at the right bottom of Fig.\ref{fig:2} shows that for long times, the wavepacket gets narrower and at time equal to 680 $T_B$ a revival occurs.} 

%\par{In Fig.~4, we plot the wavepacket dynamics for two opposite choices of $\phi$.  The results show that the dynamics are highly sensitive to drive phase, where for one phase fully coherent oscillatory transport is generated, while

\par{As shown in Fig.~\ref{fig:4}(a), a dephasing of the wavepacket is not present if we drive the system with an initial phase of $\phi=-\pi/2$. In this case coherent CBHOs emerge, which remain intact even on long time scales, however, the amplitude is reduced due to opposite polarity of drive to BOs. The vanishing of dephasing can be accredited to reduction in the amplitude and correspondingly decreased spatially-varying effects. Fig.~\ref{fig:4}(b) shows same behavior in k-space. The modulation appears immediately with a relative phase following a sine waveform which is again confined near the center of the BZ. Thus the interaction between BO and the modulation takes place at small values of quasi-momentum and accordingly the real space CBHOs are smaller in amplitude.}

\par{At this point we note that the amplitude of the CBHOs can be enhanced for modulations that modify the BO dynamics at larger absolute values of the quasi-momentum. In Fig.~\ref{fig:4}(c) we show the real space dynamics for such a case where we modulate the system with a phase of $\phi=\pi/2$. Clearly we see an increase in the amplitude of the wavepacket transport, however, we find a new kind of dynamics which correspond to a superposition of breathing and center-of-mass Bloch dynamics. In contrast to the coherent CBHOs reported above, the wavepacket distribution now spreads rapidly. The wavepacket stretches in space and we can identify two regions of unequal densities, where at one end with higher density towards the direction of force the wavepacket performs purely breathing dynamics and at the other end with lower density it undergoes anharmonic breathing. Both breathing oscillations occurring at a difference of $0.5T_B$ pass on the maximum density to a revival of the coherent BOs in just 6 Bloch periods. Furthermore, we see all three types of oscillations happening at the same time with periodic changes in density.}

\begin{figure}[t]
\hspace*{-0.3cm} 
\includegraphics[scale=0.375]{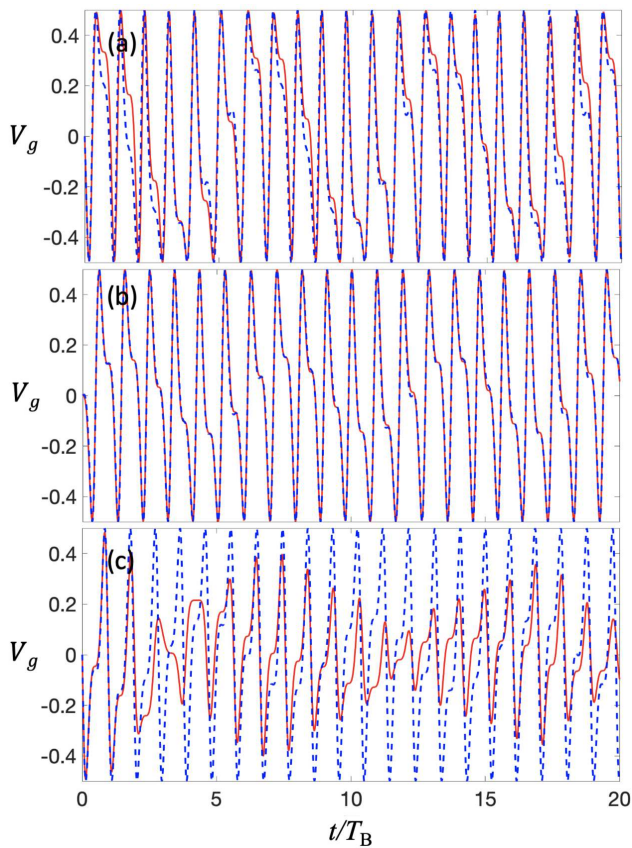}
\caption{\label{fig:epsart} Group velocity as a function of time for (a) $\phi=0$, (b) $\phi=-\pi/2$, and (c) $\phi=\pi/2$. The red line represents the result of numerical calculations, while the dashed blue line depict the dependence obtained from Eq.~\eqref{eq:6}.  For (a) the parametric values are $ \Delta F = 3.65 \times 10^{-4}~E_R $ and $\hbar \delta \omega = 5.43\times 10^{-4}~E_R $, while in (b) and (c) $\Delta F = 2.13 \times 10^{-4}~E_R$. All the other parameters are the same as used previously.} \label{fig:5}
\end{figure}
\par{The modulation appearing near the edges of the BZ is also responsible for the wavepacket spreading. This is evidenced by the wavepacket evolution in k-space. Fig.~\ref{fig:4}(d) highlights the sustained momenta near both edges which correspond to real space motion in opposite directions and by which the wavepacket spreads at times where Bragg reflections occur.  One can also interpret this in terms of a negligible cycle-averaged momentum. We see that the average momentum slowly increases due to the varying Bloch frequency and thus an asymmetric spreading is generated. Then again the wavepacket gets narrower due to phase mixing and continues following a combination of breathing and coherent dynamics.} 
\par{Note that the spreading oscillations reported here are similar in nature to the ballistic spreading regime that appears for the modulation of a constant force system at the drive phases $\phi=-\pi/2, \pi/2$. However, under the driving of a position-dependent force at $\phi= \pi/2$ we find an asymmetrical spreading motion which dies out in a few Bloch periods and which, unlike ballistic spreading, gives rise to a mix of breathing and coherent Bloch dynamics. Also contrastingly we find CBHOs at a $\phi=-\pi/2$ phase drive. }
%Taking into account the weak parabolic trap considered here, the force varies slowly in space such that we approximate it with a locally static force. With this assumption we dene the acceleration
\par{We analyze the obtained complex dynamics further by comparing them to the local acceleration model. Taking into account the weak parabolic trap considered here, the local acceleration theorem with a position- and time-dependent force of our problem, gives
\begin{equation} \label{eq:6}
	 \hbar \dot{k}_c(t) = -{2K(t)n(t)}/{d}.
\end{equation}
With regards to BO dynamics in the absence of trap modulation, within the semi-classical approach the spatial-variations during a BO have negligible effect on the dynamics  (see supplemental material). Therefore, following the harmonic oscillator-like transport during CBHOs, we approximate that the wavepacket's center in real space moves according to $n(t) = n_0 + \Delta n \sin(\delta\omega t + \gamma)$. Thus on solving Eq.~\eqref{eq:6} we get $k_c(t)$, and by perturbation method the group velocity is given as
\begin{widetext}
\begin{eqnarray} \label{eq:7}
	V_g(t) = \frac{Jd}{\hbar} \sin \!\bigg[k_od- \frac{F_{n_0}d}{\hbar} t + \frac{F_{n_0}d \alpha}{\hbar \omega_D} \;\!\big\{\cos(\omega_D t+\phi)-\cos(\phi)\big\}  + \frac{ \Delta F d}{\hbar \delta\omega}\big\{\!\cos(\delta\omega t\!+\!\gamma) \!-\!\cos(\gamma) \big\} \nonumber \\
	  	\pm \frac{ \Delta F  d \alpha}{\hbar (\omega_D \pm \delta\omega)} \;\!\big\{\sin((\omega_D \pm \delta\omega) t+(\phi\pm \gamma))-\sin(\phi \pm \gamma)\big\}  \bigg] , 
\end{eqnarray}
\end{widetext}
where $\Delta F$ is the change in force across spatial distance $\Delta n$, and $\delta \omega$ is the frequency of CBHOs. Equation~\eqref{eq:7} is plotted in Fig.~\ref{fig:5} where the analytical result is compared with the numerically-calculated dynamics of the group velocity. The parameters $\Delta n$ and $\delta \omega$ are particular to the system and we extract these from the real space dynamics shown above. Figure~\ref{fig:5}(a) shows that for $\phi=0$ the semiclassical and approximate analytical result covers the relative phase of CBHOs quite well. However, in this case the spatially-varying effects are quite significant and cannot be fully captured by the approximate model and therefore no exact match between numerics and Eq.~\eqref{eq:7} is obtained. For the case of $\phi=-\pi/2$ shown in Fig.~\ref{fig:5}(b) we achieve a very good agreement between the Eq.~\eqref{eq:7} and the numerical calculations and the rapid oscillations of the relative phase are also confirmed. Clearly, the insertion of an oscillatory function put restrictions on the relative phase such that the modulation now do not affect the entire velocity values unless $\delta \omega$ is very small. Fig.\ref{fig:5}(c) shows the breakdown of our analytical model which is due to the spreading and the multi-mode dynamics present for $\phi=\pi/2$ which are beyond the semiclassical model.}

\par{In summary, our calculations demonstrate that a position- and time-dependent force realized in ultracold atomic systems brings about an abundance of novel dynamics which are not accessible in traditional solid-state systems. The dynamics range from CBHOs, over collapse and revivals to the asymmetric spreading oscillations. These are the outcome of a phase modulation induced by the spatial variations. The k-space evolution of the wavepacket and the predictions of an approximate semiclassical model confirm our interpretations. Our studies provide a general protocol to analyze and predict the dynamics in more realistic material systems, where the lattice profile can be globally or locally inhomogeneous. Also different frequency chirps can be induced in the driving field to artificially manipulate the transport or to tailor the dynamics with unique oscillation profiles. These findings provide exciting opportunities for experimental verification with ultracold-atom experiments and may lead to new insights into the properties of quantum systems in confined geometries with potential applications in atomic diffraction, atom interferometry, and force metrology.}
\vspace{0.2cm}
\par{U.A. gratefully acknowledges support from the Deutscher Akademischer Austauschdienst (DAAD) by a doctoral research grant.}

%super-Bloch dynamics appear in the combination of the drive phase dependent transport and spreading dynamics.

%Depending upon the initial phase of the modulation field the oscillations are generated with different amplitude. The small amplitude oscillations remain coherent, while at large amplitudes the wavepacket broadens, sometimes leading to dephasing and in other conditions producing new kind of mixed ordinary and super-Bloch dynamics. These regimes depend on the drive phase which give rise to a variable detuning of  and have been linked to the relative phase appearing in the dynamics in quasimomentum space. The good agreement with an approximate semiclassical model for some initial phases confirms our interpretations. In the future, we plan to work on developing improved analytical descriptions and will consider other implementations of the variable detuning.

\end{document}